\documentclass[12pt]{revtex4}

\usepackage{amsbsy}
\usepackage{amssymb}
\usepackage{amsmath}
\usepackage{graphicx}
\usepackage{hyperref}

\usepackage{amsmath}
\usepackage{color}

\def\dw2{{\delta \omega_\xi^2}}

\newcommand{\be}{\begin{eqnarray}}
\newcommand{\ee}{\end{eqnarray}}
\newcommand{\bdm}{\begin{displaymath}}
\newcommand{\edm}{\end{displaymath}}
\newcommand{\bea}{\begin{eqnarray}}
\newcommand{\eea}{\end{eqnarray}}

\newcommand{\CDM}{{\Lambda CDM}}

\begin{document}

\title{Primordial Black Hole Remnants as Dark Energy and an Instability of De Sitter Space towards Power-Law Accelerating Expansion}

\author{Dimitris S. Kallifatides}
\email{dkallifatides@hotmail.com}


\date{\today}


\begin{abstract}
Starting from a discussion of the black hole information loss paradox, we argue that Primordial Black Hole Remnants (PBHRs) of Planck mass, should they exist, are a plausible candidate for dark energy. We call this proposal the PBHR model. We
also find an instability of de Sitter space towards a space with power-law accelerating expansion, the time dependence of the cosmological constant and the resulting form of the metric, and propose that this is the dominant instability of de Sitter space. We conjecture that a similar instability does not afflict a space sourced by PBHRs.  We find that the PBHR model satisfies the upper limits on $\beta$, the fraction of the Universe's mass in PBHs at their formation time, for a fair range of initial PBH masses. We derive information on the present day gravitational wave background due to the PBHs in our scenario.
\end{abstract}

\maketitle

\section{Introduction}
\label{sec:1}

A major recent discovery has been the acceleration of the universe expansion rate \cite{1}, subsequently corroborated by many observations \cite{2}. Many alternative theories have been put forth ever since \cite{3}, attempting to explain this acceleration of the universe expansion rate. Probably, the simplest such explanation is the $\CDM$ model, which is also in good agreement with observations with the appropriate choice of the cosmological constant \cite{4, 5}. However, one would prefer a more ``physical'' explanation of the expansion acceleration than invoking a cosmological constant.

	For the moment, let us move to another front. One of the most important contributions of Stephen Hawking has been his work on Hawking radiation, and the prediction in 1974 that black holes evaporate \cite{6}. A lot of work by him and many others since 1974 has been devoted on various aspects of this prediction. Prime examples are the concomitant black hole information loss paradox \cite{7} (or puzzle \cite{8}), and the question whether black holes leave behind any remnants after everything that can evaporate away has evaporated away.
	
In this work, we are going to argue that, if black holes leave behind remnants, primordial black hole remnants (PBHRs) of Planck mass are a plausible candidate for dark energy, which we call the PBHR model. Furthermore, using the work of Gibbons and Hawking on de Sitter space thermodynamics, we are going to argue that de Sitter space is unstable towards power-law accelerating expansion.
	In section 2 we are going to give some evidence that Planck mass PBHRs are a plausible candidate for dark energy. In section 3 we are presenting a heuristic derivation of a de Sitter space instability towards power-law accelerating expansion. In section 4 we are presenting our conjecture that a space sourced by PBHRs does not suffer from a similar instability, and thus is an observationally viable candidate for dark energy. In section 5 we show that the PBHR model satisfies the upper limits on $\beta$, the fraction of the Universe's mass in PBHs at their formation time, for a fair range of initial PBH masses, and give some information on the present day gravitational wave background due to the PBHs in our scenario. In section 6 we discuss our results.
	We are using units for which $\hbar = c = G = k = 1$, with the exception of section 5, in which we use SI units.

\section{The PBHR model}
\label{sec:2}

	The black hole information loss paradox is the following, limiting the discussion to 4D Schwarzschild black holes to keep it simple \cite{7}: As Hawking discovered, black holes emit radiation that is approximately black body radiation at a temperature 
\be
T=\frac{1}{8 \pi M}.
\ee
The main physical dependence of the black hole temperature is that it is inversely proportional to black hole mass. However, black body radiation is featureless. The only characteristic that can be discerned is the black body temperature, thus, in the case of a Schwarzschild black hole, the black hole mass. Even though there can be improvements to the black body approximation, for instance, a grey body behavior of the black hole, depending on the energy and angular momentum of the particles emitted, or, taking into account the fact that the black hole temperature increases as the black hole evaporates, both effects  introduce corrections that only depend on the black hole mass at the time of emission, as far as the black hole is involved. On the other hand, the first law of thermodynamics, as applied to a Schwarzschild black hole, 
\be
dM=T dS,
\ee
leads to the famous Bekenstein-Hawking area formula for the black hole entropy
\be
T=4 \pi M^2=\frac{A_h}{4}
\ee
where $M$ is the black hole mass, and $A_h$ is the black hole horizon area.
Thus, astrophysical black holes, like the ones discovered in 2016 \cite{9}, have a huge entropy. 
	One can then imagine the following thought experiment. One starts with a pure quantum state in the far past, that we know is going to evolve to a black hole. Since one started with a pure quantum state, the initial entropy is zero. When the black hole forms, it has a large Bekenstein-Hawking entropy. As the black hole evaporates, its Hawking radiation contains essentially no information about what formed the black hole, since it only depends on the black hole mass at the emission time, as described above. When the black hole has evaporated away, which takes a time proportional to the cube of the initial mass of the black hole, one is left with Hawking radiation from the various stages of the evaporation. This Hawking radiation has at least as much entropy as the Bekenstein-Hawking entropy of the black hole at its formation, thus the total physical system is in a mixed state, which constitutes a violation of one of the fundamental quantum mechanical laws, namely unitarity. Furthermore, since virtual black holes contribute to every quantum mechanical process, every quantum mechanical process is contaminated with unitarity violation. On the other hand, many high energy experiments have been performed over the years, for which the assumption of unitary evolution led to excellent agreement between theoretical predictions and experimental results. The fact, then, that the scientific community has been able to correctly predict experimental results of high energy experiments assuming unitarity, combined with the prediction of unitarity violation by the thought experiment above, is the essence of the black hole information loss paradox. We should add that the black hole information loss paradox generalizes to all black hole solutions.

	A lot of work has gone into attempts of resolving or ameliorating this paradox over the years. Examples are the ``black hole correspondence principle'' \cite{8}, with the ramifications analyzed by AMPS \cite{10}, the ``final state'' proposal \cite{11, 12}, and the recent work on ``soft hair'' \cite{13}. The issue has not been resolved yet \cite{14}.

	We will pick on a part of the discussion on the possible resolutions of the black hole information loss paradox, namely the suggestion that a black hole leaves behind a remnant which carries the sought information, so there is no unitarity violation. The standard response to this, is that, since black holes can have arbitrarily large mass and thus arbitrarily large entropy, there should be an infinite number of species of black hole remnants. Then, in any high energy experiment, the amplitude for remnant production would be infinite. Clearly, there has never been a report of remnant production in any high energy experiment that has already been performed, thus we are led to another paradox. In other words, if we accept that a black hole leaves behind a remnant carrying the missing information, we trade the black hole information loss paradox for the paradox involving remnant production in high energy experiments.

	We understand that the argument above is usually considered to imply that the remnant hypothesis cannot be the correct answer to the black hole information loss paradox, even though there have been attempts to save the remnant hypothesis (for instance \cite{15}). 

	We will attempt here to capitalize on the remnant hypothesis, by turning the argument on the remnant production paradox on its head. Namely, we will focus on the fact that the only real lesson from this argument is that, barring caveats \cite{15}, the assumption that a black hole leaves behind a remnant cannot be used to resolve the black hole information loss paradox. We cannot really claim that there is also the implication that a black hole does not leave behind a remnant. 

Instead, what we will keep from the discussion above is that the assumption of black hole remnants most probably implies that they have zero or little entropy. 
In particular, let us assume, for the moment, that black hole remnants carry zero entropy. This would imply that the ``final state'' proposal is correct, and, actually, this final state is a unique species unique state remnant. We will also assume that the remnant mass is $m = \kappa M_{Pl}$, where $M_{Pl} = 2.2 \times 10^{-8} {\rm kg}$ is the Planck mass, and $\kappa$ an order one numerical constant that will be left undetermined here. The assumption that the black hole remnant mass is of the order of $M_{Pl}$, is the conservative assumption.

	Clearly, we assume the theoretical, at least, existence of a new kind of particle, the zero entropy black hole remnant. An obvious question: What is this new particle good for?

	As a motivation for the rest of our discussion, let us assume that we have a collection of our remnants in a cosmological setting, namely that we have a spatially uniform density ρ of such remnants. For any particle species, a well-known equation relates its entropy density $s$, energy density $\rho$, pressure $p$, chemical potential $\mu$, number density $n$, and temperature $T$:
\be
s=\frac{\rho+p-\mu n}{T}.
\label{4}
\ee
	Since the spatial distribution of our remnants has been approximated as uniform, the collection of remnants has zero entropy, and thus zero entropy density. Applying equation (\ref{4}) to the remnants, implies that the numerator of the right-hand-side is zero. We will make the additional assumption that the chemical potential  for our remnant species vanishes, which gives
\be
\rho+p=0.
\label{5}
\ee
	Then, black hole remnants have the same equation of state as the cosmological constant.
	Of course, this is far from a proof that the existence of black hole remnants implies that they have the same equation of state as the cosmological constant. What we have actually shown is that there is a distinct possibility that they do. In particular, it is conceptually pleasing that zero entropy implies a constant energy density for the black hole remnants.
	
So far, we have adopted (\ref{5}) as the equation of state for the black hole remnants. However, we would expect corrections to (\ref{5}) depending on the UV completion of general relativity, so, instead, we will assume the equation of state
\be
w=\frac{p}{\rho}=-1+\epsilon,
\label{6}
\ee
where $\epsilon$ is a numerical constant with an absolute value much less than unity. The value and sign of $\epsilon$ will be left undetermined in this work. In particular, we cannot determine whether black hole remnants will have a quintessence-like behavior or a phantom-like behavior. In section 4 we are going to argue that $\epsilon$ is nonzero.

	On the other hand, much work has been done on primordial black holes (PBHs) \cite{16}. If they had formed in the very early universe (post-inflationary/early-radiation era), and had small enough mass, they would have evaporated by now.

	These remarks suggest the exciting, even if far-fetched, possibility that the dark energy, invoked to explain the accelerating expansion of the universe, consists of primordial black hole remnants (PBHRs) of Planck mass, obeying the equation of state (\ref{6}), and this is what we actually propose. We will call it, in short, the PBHR model. If it is true, it implies that 7/10 of the universe density consists of the stuff of quantum gravity itself, Planck mass black hole remnants. 
If true, it is an  unexpected window into quantum gravity.
It is also an extreme example of UV/IR mixing, in the sense that, if Planck mass
PBHRs exist, they  certainly belong to the UV regime, while the dark energy
has far infrared atributes, due to the extreme smallness of the dark energy density
compared to the natural (Planck) value of the vacuum energy density.

	A possible objection to our proposal is that one would expect PBHRs to behave as dark matter \cite{46}, instead of dark energy. The answer is that it is natural that Planck mass PBHRs have large quantum gravitational corrections to the classical behavior, and this may justify a dark energy behavior, instead of a dark matter behavior.

	Before working out some of the consequences of our proposal, we will interject in the next section a discussion of de Sitter space, with some novel aspects, that we are going to need. 

\section{An instability of de Sitter space}
\label{sec:3}

One of the proposals for the nature of dark energy is the cosmological constant $\Lambda$, and the concomitant $\CDM$ model, in good agreement with observations \cite{4, 5}. According to this model, the universe consists in matter and dark energy at a ratio of 3:7. For the purposes of the present note we will lump together ordinary matter and dark matter as matter or dust, since they both have vanishing pressure. We will also assume a flat ($k=0$) 4D FLRW expanding universe, with the main information being the exact time dependence of the scale factor $a(\tau)$:
\be
ds^2=-d\tau^2+a^2(\tau) \left[dr^2+r^2 \left(d\theta^2+\sin^2\theta \;d\phi^2\right)\right].
\label{7}
\ee
	Then, the Einstein equations, with some algebraic manipulation, give \cite{17}:
\be
\dot{\rho}+3(\rho+p)\frac{\dot{a}}{a} &=& 0, \label{8} \\
\dot{a}^2 &=& \frac{8 \pi \rho a^2}{3}.
\label{9}
\ee
	The behavior of the three main ingredients of the universe, namely radiation ($p=\rho/3$), matter ($p=0$), and the cosmological constant ($\rho+p=0$), is given by the following equations, respectively:
\begin{itemize}
\item	For radiation:
\be
\rho a^4 &=& {\rm const} \label{10}\\
a(\tau) &=& (4 C)^{1/4} \tau^{1/2}, \label{11}
\ee
where $C=8 \pi \rho a^4/3$.
\item For matter:
\be
\rho a^3 &=& {\rm const} \label{12} \\
a(\tau) &=& \left( \frac{ 9 C'}{4} \right)^{1/3} \tau^{2/3}, \label{13}
\ee
where $C'=8 \pi \rho a^3/3$.
\item For the cosmological constant:
\be
\rho  &=& {\rm const} \label{14}\\
a(\tau) &=& a_0 e^{\sqrt{C''} (\tau-\tau_0)}, \label{15}
\ee
where $C''=8 \pi \rho /3$ and $a_0=a(\tau_0)$.
\end{itemize}

	We will focus on the cosmological constant. Clearly, equations (\ref{14}) and (\ref{15}) are only valid in the case of stable de Sitter space. However, various instabilities of de Sitter space have been discussed \cite{18}-\cite{42}, which would modify equations (\ref{14}) and (\ref{15}).

	It has been proposed \cite{18}-\cite{25} that the instability of de Sitter space is toward lower values of the cosmological constant $\Lambda$. Mottola \cite{23}, in particular, argued that, since the specific dependence of the de Sitter space entropy \cite{43} on the cosmological constant $\Lambda$,
\be
S_{dS}=\frac{3 \pi}{\Lambda},
\label{16}
\ee
showed that the de Sitter space entropy increases when $\Lambda$ decreases, there is an instability of de Sitter space toward $\Lambda=0$.

	Here, we will use a heuristic way to quantify this argument, and suggest that this instability of de Sitter space toward  $\Lambda=0$ is the dominant instability of de Sitter space. As a by-product, we will get a new equation of state for the cosmological constant, and equations replacing (\ref{14}) and (\ref{15}).

	First, it is common knowledge that Einstein equations with a cosmological constant  $\Lambda$ are equivalent to the ``ordinary'' Einstein equations with a perfect fluid energy-momentum tensor on the right-hand-side, for which
\be
\rho+p&=&0 \label{17}\\
\Lambda &=& 8 \pi \rho.
\label{18}
\ee
	On the other hand, de Sitter space has entropy, given by equation (\ref{16}), and temperature \cite{43}, given by
\be
T_{dS}=\frac{1}{2\pi} \sqrt{\frac{\Lambda}{3}}.
\label{19}
\ee
	We have already mentioned equation (\ref{4}), valid for any particle species. We would like to apply equation (\ref{4}) to de Sitter space, assuming a vanishing chemical potential, and the sum $\rho+p$ as an unknown. First, we need to compute the entropy density. To do that, we divide the entropy (\ref{16}) by the appropriate volume. Now, what is the appropriate volume to divide with? Since the entropy (\ref{16}) is proportional to the area of the horizon of the static patch,
\be
ds^2=-\left(1-\frac{r^2}{l^2}\right) dt^2+\left(1-\frac{r^2}{l^2}\right)^{-1} dr^2
+r^2(d\theta^2+\sin^2\theta \; d\phi^2),
\label{20}
\ee
with $0\leq r < l$ and $\Lambda=3/l^2$,
the volume related to the entropy density calculation is the volume of a constant time hypersurface of the static patch:
\be
V=\pi^2 l^3.
\label{21}
\ee
This gives a de Sitter space entropy density
\be
s=\frac{S_{dS}}{V}=\frac{1}{\pi l}=\frac{1}{\pi}\sqrt{\frac{\Lambda}{3}}=\sqrt{\frac{8 \rho}{3\pi}},
\label{22}
\ee 
and, using 
(\ref{19}), an equation of state for the ``unstable'' de Sitter fluid
\be
w=\frac{p}{\rho}=\frac{4}{3\pi}-1 \simeq -0.58.
\label{23}
\ee
Feeding then back (\ref{23}) to equations (\ref{8}) and (\ref{9}), we derive the following equations for ``unstable'' de Sitter space:
\be
\rho a^{4/\pi} &=&{\rm const} \label{24} \\
a(\tau) &=& \left(\frac{4 C_3}{\pi^2}\right)^{\pi/4} \tau^{\pi/2},
\label{25}
\ee
with $C_3=8 \pi \rho a^{4/\pi}/3$.
These replace (\ref{14}) and (\ref{15}), which are valid for stable de Sitter space. Using 
(\ref{18}), (\ref{24}) implies
\be
\Lambda=\frac{3 \pi^2}{4} \frac{1}{\tau^2}.
\label{26}
\ee
Thus the cosmological constant is {\it not} a constant, but instead is completely determined as a function of time, which is normal for power-law expansion. 
Clearly, (\ref{26}) quantifies an instability of de Sitter space toward $\Lambda=0$, even though the time to reach the value $\Lambda=0$ diverges. Specifically, (\ref{25}) has a power-law accelerating expansion behavior with the power $\pi/2$. We propose that this is really the dominant instability of de Sitter space. 

	To our knowledge, in this work it is the first time that the work of Gibbons and Hawking on de Sitter space thermodynamics \cite{43} is employed to derive definite results on the de Sitter space instability.

	In the process of deriving equations (\ref{23}–-\ref{26}), we used equation (\ref{4}), even though the de Sitter space entropy (\ref{16}) is nonextensive. Indeed, we are going to make the assumption that equation (\ref{4}) is valid, even if the entropy is nonextensive. This way, for de Sitter space we found a ``holographic back reaction of thermodynamics on the metric''. We call the back reaction holographic, since the de Sitter space entropy (\ref{16}) is proportional to the area of the horizon of the causal patch.

	Furthermore, we called it a heuristic derivation, since most of the literature on de Sitter space instabilities tackles the issue the ``hard way'', by discussing the fluctuations of quantum fields and particle production on de Sitter space. What we are doing here instead, is bypass the usual analysis, by using equation (\ref{4}) and the thermodynamics of de Sitter space.
	
One aspect of the instability we found, is that for an expanding universe the cosmological constant diminishes with time, as described by (\ref{24}). This ameliorates the problem of the smallness of the cosmological constant by several orders of magnitude. In the process of deriving (\ref{24}), we assumed a vanishing chemical potential for the de Sitter ``fluid''. If, instead, we assume a positive chemical potential for the de Sitter ``fluid'', this will lead to a reduction of the cosmological constant at a faster pace than in (\ref{24}), and a more decisive amelioration of the problem of the smallness of the cosmological constant.

\section{Dark energy consists of PBHRs}
\label{sec:4}

In the previous section, we found an instability of de Sitter space, as described by equations (\ref{24} -- \ref{26}), and proposed that this is the dominant instability of de Sitter space. Aspects of this instability is that for the de Sitter ``fluid'' $w$ is about $-0.58$, and that for an expanding universe the cosmological constant diminishes with time, as described by (\ref{24}). We may attempt to modify the $\CDM$ model, adopting the behavior (\ref{24}) for the cosmological constant, with a present value of $7 \times 10^{-27} \;{\rm kg}/{\rm m}^3$ for the dark energy density, with a universe consisting in matter and dark energy at a ratio of 3:7. Then, using equation  (\ref{24}), one finds that, when the universe temperature was $6 \;{\rm K}$, the dark energy density was $1.9 \times 10^{-26} \;{\rm kg}/{\rm m}^3$, bigger than the present matter density by a factor of about 6, more in line with \cite{44} and \cite{45}.

However, the scenario of the previous paragraph is contrary to observations \cite{5}. According to observations \cite{5}, $w = p/\rho$ is close to -1 for the dark energy, with the dark energy density being, at least approximately, constant.  Thus, given the de Sitter space instability we found, this is evidence against the hypothesis that dark energy is the cosmological constant.
A logical alternative is that the PBHR model of section 2 has the same instability. In that case, one might expect the PBHR density in an expanding universe to diminish with time, as described by (\ref{24}). However, if additional black holes produce new remnants with the passage of time, it is conceivable that the PBHR density diminishes with time at a slower rate than that given by (\ref{24}), or not at all. 

This last scenario has at least two drawbacks:
\begin{itemize}
\item	It takes a considerable degree of fine tuning to have at the same time an unstable PBHR model, and an approximately constant PBHR density, due to the production of new remnants by additional black holes.
\item	This scenario is not observationally viable, because it leads to an enormous $\gamma$ radiation background.
\end{itemize}

What we are going to suggest instead, is that the thermodynamics of a space sourced by Planck mass black hole remnants is sufficiently different from de Sitter space, so that the former has no instability. A necessary condition for this is that the PBHR model satisfies equation (\ref{6}) with a nonvanishing $\epsilon$. We are also going to assume, like we did in section 2, that the absolute value of 
$\epsilon$ is much smaller than 1,  thus $w$ is close to -1 for the PBHR model. Then, the PBHR model is an observationally viable candidate for dark energy.
In addition, we assume that the cosmological constant vanishes by some other mechanism (for instance \cite{50}).

In the following section, we are going to provide some evidence in favor of the PBHR model, finding that it satisfies the upper limits on $\beta$, the fraction of the Universe's mass in PBHs at their formation time, for a fair range of initial PBH masses. We are also going to provide some features of the gravitational wave background generated by these PBHs.

\section{The fraction of the Universe's mass in PBHs at their formation time}
\label{sec:5}

In this section, we are going to sketch a calculation, according to which, the PBHR  model satisfies the upper limits on $\beta$, the fraction of the Universe's mass in PBHs at their formation time, for a fair range of initial PBH masses.
The $\CDM$ model is in good agreement with observations \cite{4, 5}, for the value 
$7 \times 10 ^{-27} \;{\rm kg}/{\rm m}^3$ of the dark energy density, approximately. For the proposal that the dark energy consists of primordial black hole remnants (PBHRs), the PBHR model, and assuming a mass $m = \kappa M_{Pl}$ for the black hole remnant, where $M_{Pl} = 2.2 \times 10^{-8} \;{\rm kg}$ is the Planck mass, and $\kappa$ an order one numerical constant, then the present number density of PBHRs is parametrized by $\kappa$, as
\be
n=\frac{3}{\kappa} 10^{-19}.
\label{27}
\ee
	One would like to determine, among other physical quantities, the masses of the primordial black holes (PBHs) when formed, the time at which they formed (formation time), the approximate time at which they had evaporated leaving behind PBHRs, the fraction $\beta$ of the Universe's mass in PBHs at their formation time, and so on. 

	For simplicity, we will assume that all PBHs are Schwarzschild black holes of the same mass 
$M$ at their formation time, and they all form at the same time $t_f$. We will also ignore the possibility of coalescence between the PBHs.

	To start with, given that the temperature of a Schwarzschild black hole is given by \cite{7}
\be
T=\frac{M_{Pl}^2 c^2}{8 \pi k M},
\label{28}
\ee
and substituting the values of the physical constants, we parametrize the black hole temperature in terms of its mass $M$ in SI units, as
\be
T=1.3 \times 10^{23} M^{-1}.
\label{29}
\ee
	
A reasonable approximation for the temperature of the Universe at time $t = 1 \;{\rm s}$, is $T_{U} = 10^{10} {\rm K}$. Then, assuming that these processes occur in the early radiation era, the Universe temperature $T_{U}$ at time $t$ can be parametrized in terms of $t$, as
\be
T_{U}=10^{10} t^{-1/2}.
\label{30}
\ee
A very probable value $t_{f,p}$ for the PBH formation time $t_f$ is when the
universe temperature is (in Planck units) equal to the scale of density of the collapsing region:
\be
t_{f,p}=2 \times 10^{-36} M.
\label{31}
\ee
	Given that today 7/10 of the Universe is dark energy, with a density of approximately 
$7 \times 10^{-27} {\rm kg}/{\rm m}^3$, the 3/10 of matter has an approximate density of 
$3 \times 10^{-27} {\rm kg}/{\rm m}^3$. We lump ordinary matter and dark matter together, since they have the same equation of state.
For the purposes of our sketchy calculation, we will assume that at Universe temperature 
$2.7 \times 10^4 {\rm K}$, radiation density and matter density are equal. Then, using the present universe temperature of about $2.7 {\rm K}$, and equation (\ref{12}), we find that the radiation and matter density equal each other at a value of about $3 \times 10^{-15} {\rm kg}/{\rm m}^3$.
Setting then the Universe temperature at $10^{10} {\rm K}$ at time $1 {\rm s}$, we find that at time $1 {\rm s}$ the radiation density is about $6 \times 10^7 {\rm kg}/{\rm m}^3$. Using (\ref{30}), we parametrize the radiation density at the time of formation of the PBHs, as
\be
\rho_{rf}=6 \times 10^7 t_f^{-2}.
\label{32}
\ee

	In order to estimate the evaporation time of the PBHs, we will assume that they are mini ones, namely that $k T$ is at least $200 {\rm GeV}$ at formation time. We will actually find that only for such mini black holes the upper limits on $\beta$ \cite{46} are satisfied. This implies that, due to Hawking radiation of the PBHs, all known fundamental particles are emitted in the relativistic regime. In \cite{47} a table is given with the grey body factors for all spins. As a result, we get the following equation for the evaporation time of a PBH with $k T$ at least 
$200 {\rm GeV}$ at formation:
\be
t_{ev}=4 \times 10^{-19} M^3.
\label{33}
\ee
$k T$ of at least $200 {\rm GeV}$ corresponds to a maximum mass of about 
$6 \times 10^7 {\rm kg}$. Combining with equation (\ref{33}), we get an
evaporation time $t_{ev}$ 
$9 \times 10^4 \; {\rm s}$. We will assume a formation time of less than $10^{-4} {\rm s}$, and, thus, the evaporation of the PBHs we are considering ends in the early radiation era.
	Combining equations (\ref{30}) and (\ref{33}), we get the Universe temperature at the time of complete evaporation:
\be
T_{ev}=10^{10} (t_f+4 \times 10^{-19} M^3)^{-1/2}.
\label{34}
\ee
	Combining (\ref{33}) with (\ref{6}), we parametrize the PBHR density at the time of the end of evaporation, in terms of the initial black hole mass and the formation time, as
\be
\rho_{PBHR}=7 \times 10^{-27} 
\left[
\frac{10^{10} (t_f+4 \times 10^{-19} M^3)^{-1/2}}{2.7}
\right]^{3 \epsilon},
\label{35}
\ee
and, the PBHR number density at the time of the end of evaporation, as
\be
n_{PBHR}=\frac{3}{\kappa} \times 10^{-19} 
\left[
\frac{10^{10} (t_f+4 \times 10^{-19} M^3)^{-1/2}}{2.7}
\right]^{3 \epsilon}.
\label{36}
\ee
	Identifying this with the PBH number density at the end of evaporation, and using
\be
n_{PBH} a^3={\rm const}
\label{37}
\ee
for the PBH number density, we get the PBH number density at formation,
\be
n_{PBH,f}=
\frac{3}{\kappa} \times 10^{-19}  
\left[
\frac{10^{10} }{2.7}
\right]^{3 \epsilon}
 \frac{\left( t_f+4 \times 10^{-19} M^3 \right)^{3(1-\epsilon)/2}}{t_f^{3/2}},
\label{38}
\ee
and the PBH density at formation:
\be
\rho_{PBH,f}=
\frac{3}{\kappa} \times 10^{-19}   M 
\left[
\frac{10^{10} }{2.7}
\right]^{3 \epsilon}
\frac{\left( t_f+4 \times 10^{-19} M^3 \right)^{3(1-\epsilon)/2}}{t_f^{3/2}}.
\label{39}
\ee
	Dividing by (\ref{32}) we get $\beta$, the fraction of the Universe in PBHs at their formation time:
\be 
\beta \simeq \frac{\rho_{PBH,f}}{\rho_{rf}}=
\frac{5}{\kappa} \times 10^{-27}   M
\left[
\frac{10^{10} }{2.7}
\right]^{3 \epsilon}
t_f^{1/2}
\left( t_f+4 \times 10^{-19} M^3 \right)^{3(1-\epsilon)/2}.
\label{40}
\ee
	The values for $\beta$ that we get from (\ref{40}) easily satisfy the upper limits on 
$\beta$ \cite{46} for an initial PBH mass of $6 \times 10^7 \;{\rm kg}$ or less. 
For instance, setting $\kappa = 1$, $\epsilon = 0.05$, $t_f$ given by (\ref{31}),
$M = 6 \times 10^7 \;{\rm kg}$, Eq. (\ref{40}) gives $\beta = 8.6 \times 10^{-25}$, 
near the limit of the range of allowed values. 


	We could also determine the main features of the present day gravitational wave background, resulting from the formation of these PBHs. The peak frequency of GWs today is given by \cite{48}
\be
f_{GW}=0.03 \;{\rm Hz}
\left( \frac{M}{10^{17} \;{\rm kg}}\right)^{-1/2}
\left( \frac{g_{\star P}}
{106.75}\right)^{-1/12},
\label{41}
\ee
where $M$ is the PBH mass at formation, and $g_{\star P}$ is the effective number of relativistic degrees of freedom when the comoving scale $k_p^{-1}$ entered the Hubble radius, with $k_p$ the wavenumber of the peak scale of fluctuations causing the PBHs. Eq. (\ref{41}) is valid when the PBH formation time $t_f$ has the value 
given by (\ref{31}).	Approximating $g_{\star P}$ with 106.75, we get:
\be
f_{GW}=9.5 \times 10^6 M^{-1/2}.
\label{42}
\ee
	With an upper value of $6 \times 10^7 \;{\rm kg}$ for $M$, this gives a minimum peak frequency of about $1.2 \times 10^3 \;{\rm Hz}$.
	\cite{49} gives an alternative relation for the peak frequency of the present day GW spectrum:
\be
M \simeq 50 \gamma \left(\frac{10^{-9} \;{\rm Hz}}{f_{GW}} \right)^2 M_{\odot}.
\label{43}
\ee
Here $\gamma<1$ accounts for the efficiency of the collapse.	
With an approximate value of 0.2 for $\gamma$ \cite{49}, this gives:
\be
f_{GW}= 4.5 \times 10^{6} M^{-1/2}.
\label{44}
\ee
	With an upper value of $6 \times 10^7 \;{\rm kg}$ for $M$, this gives an alternative minimum peak frequency of $6 \times 10^2 \;{\rm Hz}$.
	Another important relation regarding the GW spectrum is \cite{49}
\be
\beta=0.1 \exp\left(-\frac{\Psi_c^2}{2 A^2} \right),
\label{45}
\ee
where $\Psi_c$ is the threshold value of scalar perturbations for PBH formation, and $A^2$ is the $({\rm amplitude})^2 \times ln({\rm peak width})$. A typical value for $\Psi_c$ is $1/2$. The gravitational waves produced are parametrized \cite{51} by their energy-density
spectra, $\Omega(f)=\frac{1}{\rho_c} \frac{d\rho_{GW}}{d\ln f}$, 
where $d\rho_{GW}$
is the energy density in gravitational waves per logarithmic frequency interval $d\ln f$,
and $\rho_c=\frac{3H_0^2 c^2}{8 \pi G}$ is the closure energy density of the Universe.
The value of $\Omega(f)$ at the peak frequency today, $f_{GW}$, is denoted by
$A_{GW}$, and is called the amplitude of the present day spectrum at the peak frequency
\cite{48}. Then:
\be
A_{GW}=6 \times 10^{-8}
\left( \frac{g_{\star P}}{106.75} \right)^{-1/3}
\left( \frac{A^2}{10^{-2}}\right)^2.
\label{46}
\ee
	For $\beta = 8.6 \times 10^{-25}$, as in the $M = 6 \times 10^7 \;{\rm kg}$ 
case, and approximating again $g_{\star P}$ with 106.75, we get an amplitude of 
$3 \times  10^{-9}$ at the peak frequency. This result is consistent with the experimental limits on the energy-density of the gravitational wave background with tensor polarizations
\cite{51}.

	So far, we have assumed that a space sourced by PBHRs does not suffer from a similar instability to that of section 3, keeping in line with observations.  If, instead, we assume that the PBHRs source a de Sitter space suffering from the same instability we found in section 3, then 
$\beta$ is given by:
\be
\beta \simeq \frac{\rho_{PBH,f}}{\rho_{rf}}=
\frac{7.5}{\kappa}\times 10^{-15} M t_f^{1/2}
(t_f+4 \times 10^{-19} M^3)^{\frac{3}{2}-\frac{2}{\pi}}.
\label{47}
\ee
In order to obtain a value for $\beta$ in the allowed range \cite{46},
$M$ must be at most $10^7\;{\rm kg}$. For $M = 10^7 \;{\rm kg}$,
$\kappa=1$, and $t_f$ given by Eq.~(\ref{31}), Eq.~(\ref{47}) gives 
$\beta = 5.5 \times 10^{-20}$. For these values of $M$ and $\beta$, 
(\ref{42}, \ref{45}, and \ref{46}) give a minimum peak frequency of the GW spectrum today at 
$3 \times 10^3 \;{\rm Hz}$, and an amplitude of $5 \times 10^{-9}$ 
at the peak frequency. This result for the amplitude is also consistent
with the experimental limits on the energy-density of the gravitational wave background with tensor polarizations \cite{51}. Instead, (\ref{44}) gives a minimum peak frequency of $1.4 \times 10^3 \;{\rm Hz}$.

\section{Discussion}
\label{sec:6}
        
	In this work, we argued that Planck mass PBHRs have the equation of state (\ref{6}), with 
$\epsilon$ a so far undetermined nonzero constant, with an absolute value much smaller than unity. The black hole remnant mass is parametrized as $m = \kappa M_{Pl}$, where 
$M_{Pl}$ is the Planck mass, and $\kappa$ an order one numerical constant, which is also undetermined for now. We also argued for an instability of de Sitter space towards power-law accelerating expansion, as described by equations (\ref{23} - \ref{26}). Given that observations favor a $w = p/\rho$ close to -1, and a dark energy density almost or exactly constant, our instability results imply that dark energy cannot be the cosmological constant. Instead, we proposed that dark energy consists of PBHRs satisfying the equation of state (\ref{6}), as above, which we called the PBHR model. The PBHR model is an extreme case of UV/IR mixing.

	We showed that the PBHR model  satisfies the upper limits on $\beta$, the fraction of the Universe's mass in PBHs at their formation time, for mini PBHs with an upper initial mass of about $6 \times 10^7 \;{\rm kg}$. We also derived some information on the present day gravitational wave background due to the PBHs in this model. We should add that the PBHR model suffers from no coincidence problems.
	If, for the sake of the argument, we assume that a space sourced by PBHRs has the same de Sitter space instability we found, the upper limits on $\beta$ are satisfied for mini black holes with an initial mass of at most $10^7 \;{\rm kg}$, and we gave the corresponding present day gravitational wave background information.

	There are many open questions regarding the PBHR model:
\begin{itemize}
\item
Find the contribution of the gravitational wave background, due to the PBHs in this model, on the B mode of the CMB.
\item Constrain the model further by taking into account the various mechanisms of PBH production.
\item Explore how the results are affected, if we relax some simplifying assumptions.
\item Explore the question whether the de Sitter space instability found here has any bearing on inflation, especially the ``graceful exit'' problem, and eternal inflation.
\item Relate the de Sitter space instability to the literature on the fluctuations of quantum fields and particle production on de Sitter space.
\item Generalize the de Sitter space instability to different dimensionalities, and inquire whether a similar ``backreaction of thermodynamics on the metric'', based on equation (\ref{4}), happens in other spacetimes.
\item Explore what UV completion of general relativity can justify the equation of state (\ref{6}) for the BHRs, and what  information, for instance on the values of $\epsilon$ and $\kappa$, one can derive that way.
\item Investigate whether the de Sitter space instability is related to the difficulty
in finding stringy de Sitter vacua (see the discussion in \cite{52}). 
\end{itemize}
We believe that these issues deserve further investigation.

\section*{Acknowledgements}

	I am very grateful to Theocharis Apostolatos and Erik Katsavounidis for valuable discussions.


\begin{thebibliography}{9}

\bibitem{1} Perlmutter, S. et al., Astrophys. J. 517, 565–586 (1999); Riess, A. G. et al., Astron. J. 116, 1009–1038 (1998).
\bibitem{2} Miller, A. D. et al., Astrophys. J. Lett. 524, L1–L4 (1999); de Bernardis, P. et al., Nature 404, 955–959 (2000); Hanany, S. et al., Astrophys. J. Lett. 545, L5–L9 (2000); Halverson, N. W. et al., Astrophys. J. 568, 38–45 (2002); Mason, B. S. et al., astro-ph/0205384 (2002); Benoit, A. et al., astro-ph/0210306 (2002); Spergel, D. N. et al., astro-ph/0302209 (2003); Page, L. et al., astro-ph/0302220 (2003).
\bibitem{3} E. J. Copeland, M. Sami and S. Tsujikawa, Int. J. Mod. Phys. D 15, 1753 (2006) doi:10.1142/S021827180600942X [hep-th/0603057].
\bibitem{4} Planck Collaboration, Astronomy and Astrophysics. 571: A1, arXiv:1303.5062.
\bibitem{5} Paul J. Steinhardt; Neil Turok (2006), Science. 312 (5777): 1180–1183, arXiv:astro-ph/0605173; D. M. Scolnic, et al., astro-ph/1710.00845.
\bibitem{6} S. W. Hawking, Nature 248, 30 (1974).
\bibitem{7} S. W. Hawking, Phys. Rev. D 14, 2460 (1976).
\bibitem{8} L. Susskind, Phys. Rev. Lett. 71: 2367-2368, 1993.
\bibitem{9} LIGO Scientific Collaboration and Virgo Collaboration,  Physical Review Letters 116 (6): 061102, arXiv:1602.03837.
\bibitem{10} Ahmed Almheiri, Donald Marolf, Joseph Polchinski, and James Sully. Black Holes: Complementarity or Firewalls? JHEP, 02:062, 2013.
\bibitem{11} Gary T. Horowitz, Juan Maldacena, JHEP 0402:008 (2004), arXiv:hep-th/0310281.
\bibitem{12} Seth Lloyd, John Preskill, JHEP 08 (2014) 126, arXiv:hep-th/1308.4209.
\bibitem{13} Stephen W. Hawking, Malcolm J. Perry and Andrew Strominger, Phys. Rev. Lett. 116, 231301 (2016).
\bibitem{14} Sasha Haco, Stephen W. Hawking, Malcolm J. Perry and Andrew Strominger, hep-th/1810.01847.
\bibitem{15} Marios Christodoulou, Tommaso De Lorenzo, Phys. Rev. D94, 104002 (2016).
\bibitem{16} Hawking, S. (1971), Mon. Not. R. Astron. Soc. 152: 75.
\bibitem{17} Robert M. Wald, “General Relativity”, The University of Chicago Press (1984).
\bibitem{18} A. M. Polyakov, Sov. Phys. Usp. 25 (1982) 187.
\bibitem{19} I. Antoniadis and E. Mottola, Phys. Rev. D 45 (1992) 2013.
\bibitem{20} N. C. Tsamis and R. P. Woodard, Phys. Lett. B 301 (1993) 351.
\bibitem{21} L. H. Ford, Phys. Rev. D 31 (1985) 710.
\bibitem{22} E. Mottola, Phys. Rev. D 31 (1985) 754.
\bibitem{23} E. Mottola, Phys. Rev. D 33, 6 (1986) 1616.
\bibitem{24} I. Antoniadis, J. Iliopoulos and T. N. Tomaras, Phys. Rev. Lett. 56 (1986) 1319.
\bibitem{25} P. Mazur and E. Mottola, Nucl. Phys. B 278 (1986) 694.
\bibitem{26} V. K. Onemli and R. P. Woodard, Class. Quant. Grav. 19 (2002) 4607; Phys. Rev. D 70 (2004) 107301.
\bibitem{27} I. Antoniadis, P. O. Mazur and E. Mottola, New J. Phys. 9 (2007) 11.
\bibitem{28} A. M. Polyakov, Nucl. Phys. B 797 (2008) 199; Nucl. Phys. B 834 (2010) 316; arXiv:1209.4135 [hep-th].
\bibitem{29} D. Krotov and A. M. Polyakov, Nucl. Phys. B 849 (2011) 410.
\bibitem{30} D. Marolf and I. A. Morrison, Phys. Rev. D 82 (2010) 105032; Phys. Rev. D 84 (2011) 044040.
\bibitem{31} S. Hollands, Commun. Math. Phys. 319 (2013) 1.
\bibitem{32} D. Boyanovsky and R. Holman, JHEP 1105 (2011) 047.
\bibitem{33} N. C. Tsamis and R. P. Woodard, Int. J. Mod. Phys. D
20 (2011) 2847.
\bibitem{34} P. R. Anderson and E. Mottola, Phys. Rev. D 89 (2014) 104038; Phys. Rev. D 89 (2014) 104039.
\bibitem{35} E. T. Akhmedov, Phys. Rev. D 87 (2013) 044049.
\bibitem{36} A. Kaya, Phys. Rev. D 87 (2013) 123501.
\bibitem{37} E. T. Akhmedov, U. Moschella, K. E. Pavlenko and F. K. Popov, Phys. Rev. D 96 (2017) 025002.
\bibitem{38} T. Markkanen, Eur. Phys. J. C 78 (2018) 97.
\bibitem{39} P. R. Anderson, E. Mottola and D. H. Sanders, Phys. Rev. D 97 (2018) 065016.
\bibitem{40} S. P. Miao, N. C. Tsamis and R. P. Woodard, Phys. Rev. D 95 (2017) 125008.
\bibitem{41} E. T. Akhmedov, F. K. Popov and V. M. Slepukhin, Phys. Rev. D 88 (2013) 024021.
\bibitem{42} G. Moreau, J. Serreau, hep-th/1808.00338.
\bibitem{43} G. W. Gibbons and S. W. Hawking, Phys. Rev. D 15, 2738 (1977).
\bibitem{44} S. Weinberg, Phys. Rev. Lett. 59 (22), 1987.
\bibitem{45} A. Vilenkin (2006): “Many worlds in one: The Search For Other Universes”, New York: Hill and Wang.
\bibitem{46} B. J. Carr, K. Kohri, Y. Sendouda, J. Yokoyama, Phys. Rev. D81: 104019, 2010.
\bibitem{47} O. Lennon, J. March-Russell, R. Petrossian-Byrne, H. Tillim, hep-ph/1712.07664.
\bibitem{48} R. Saito, J. Yokoyama, astro-ph/0812.4339.
\bibitem{49} N. Bartolo, et al., astro-ph/1810.12218.
\bibitem{50} J. R. Klauder, hep-th/1803.05823.
\bibitem{51} LIGO Scientific Collaboration and Virgo Collaboration, gr-qc/1903.02886.
\bibitem{52} Ulf Danielsson, hep-th/1809.04512.

\end{thebibliography}
\end{document}